 \documentclass[11pt,a4paper,groupedaddress,preprintnumbers,nofootinbib,floatfix, showpacs, notitlepage
]{revtex4-1}

\usepackage[top=3.2cm, bottom=2.5cm, left=2.3cm, right=2.3cm]{geometry}
\usepackage[english]{babel}
\usepackage{amsmath,amssymb,amsfonts, dsfont}
\usepackage{graphicx}
\usepackage{color}
\usepackage{hyperref}
\usepackage{subfigure}
\usepackage{dcolumn}
\usepackage{bm}
\usepackage{float}
\usepackage{mathtools}
\usepackage{bbm}
\usepackage{pxfonts}
\usepackage[vcentermath]{youngtab}
\usepackage[all]{xy}
\usepackage{pstricks}
\usepackage{accents}
\usepackage{cases}
\usepackage{comment}
\usepackage{placeins}
\usepackage{xspace}
\usepackage{cancel} 
\usepackage{slashed}
\usepackage{hyperref}
\hypersetup{
    bookmarks=true,         
    unicode=false,          
    pdftoolbar=true,        
    pdfmenubar=true,        
    pdffitwindow=false,     
    pdfstartview={FitH},    
    pdftitle={My title},    
    pdfauthor={Author},     
    pdfsubject={Subject},   
    pdfcreator={Creator},   
    pdfproducer={Producer}, 
    pdfkeywords={keyword1, key2, key3}, 
    pdfnewwindow=true,      
    colorlinks=true,       
    linkcolor=purple,          
    citecolor=byz,        
    filecolor=orange,      
    urlcolor=blue           
}

\newcommand{\betabeta}{\mbox{$(\beta \beta)_{0 \nu}  $}}
\newcommand{\be}{\begin{equation}}
\newcommand{\ee}{\end{equation}}

\definecolor{byz}{rgb}{0.44, 0.16, 0.39}

\begin{document}
\vspace*{1cm}
\title{\textcolor{byz}{A Radiatively induced Elementary Goldstone Higgs in SU(4)/Sp(4)}}

\author{A. Meroni}

\affiliation{ CP$^3$-Origins, University of Southern Denmark, Campusvej 55, DK-5230 Odense M, Denmark.}

\begin{abstract}  
 Using a  $SU(4)\to Sp(4)$ pattern of chiral symmetry breaking, we investigate the pseudo-Goldstone  nature of the Higgs boson in an elementary realisation that at the same time provides an ultraviolet completion. The renormalizability of the model together with the perturbative corrections determine dynamically the direction of vacuum of the theory  and the corresponding  Higgs chiral symmetry breaking scale  $ f \simeq 14~$TeV. The Higgs boson is radiatively generated and the scalar mass spectrum, together with a second massive Higgs boson, lie in the multi-TeV range.
 \end{abstract}
 
 \maketitle
\section{Introduction}

The discovery   on the 4th July of 2012, by ATLAS and CMS, of a new boson
with a mass approximately of  125 GeV, decaying into $\gamma \gamma$, $WW$ and $ZZ$ bosons \cite{Agashe:2014kda}, is
with no doubts crucial in order to  unravel  the long standing problem of the origin of the mass.\\ 
At the same time, the Standard Model (SM)  per se suffers from a number of theoretical and phenomenological weaknesses, for instance:
i) the lack  of  dynamical motivation for the origin of the spontaneous symmetry breaking 
ii) the absence of a mechanism stabilising the electroweak scale against quantum corrections (the so called \emph{hierarchy problem}) 
iii) the absence of  absolute vacuum stability and 
iv) an explanation of the baryon asymmetry of the universe (BAU).\\
More importantly,  we know   that the SM
cannot be the ultimate theory, since does not account the so called ``elusive'' sector: the neutrinos and dark matter particles are  not properly included in the SM, since we cannot describe their mass. \\
In connection with this, we still do not have: 
i) a solution to the flavour puzzle, namely an explanation of the mass differences in 
the  spectrum and of the quark and lepton mixing patterns,  
 and ii) we do not have any indication about a possible
connection between non zero neutrino masses and  symmetries that could predict the lepton mixing (very different from the quark mixing).  
Moreover, the nature of the three light active neutrinos $\nu_j$ ($j=1,2,3$) with definite mass $m_j$ is unknown.  
Neutrinos can be
Dirac fermions if particle interactions conserve
some additive lepton number, e.g., the total
lepton charge $L = L_e + L_{\mu} + L_{\tau}$.
If $\nu_j$  are found to be Majorana fermions,   no lepton charge can be conserved \cite{Bilenky:2001rz}\cite{Bilenky:1987ty}. The only feasible experiment that can unveil the nature of massive neutrinos is neutrinoless double beta, \betabeta decay (see e.g. \cite{Rodejohann:2011mu} for a review).
An attractive  explanation of the Majorana nature of massive neutrinos is provided 
by the See-Saw mechanism \cite{seesaw}, which not only gives an explanation of the
smallness of neutrino masses, through the existence of heavier fermionic singlets, but also gives a explanation to the 
observed BAU, through the leptogenesis theory \cite{LeptoG}.
\\
Despite the discovery of a new, subatomic particle, at present it is not yet clear if the mechanism 
observed in the experiments is the one originally envisioned.
Following the symmetry principle and motivated by the issues of the SM raised above, it is therefore appealing to consider a  larger ---unified--- symmetry 
at higher energy scale, embedding the SM one.
This hypothesis  implies the existence of   a different Higgs sector and this would lead to a complete new phenomenology   
via the Yukawa couplings (in particular one expects sizeable effects from the heaviest fermion 
of the SM, the top quark) through the interactions with new charged and neutral scalars.
Such a scenario can be realised by introducing  the  Higgs boson  as a \emph{fundamental} pseudo Nambu-Goldstone boson (pNGB)
of a new theory,  which is supposed to be invariant at a scale $f\gg v_{ew}$, under a global symmetry
$G$ which is spontaneously broken to a stability group $H$.  
 Models predicting the Higgs as a pNGB are relevant alternatives to the SM. One can 
achieve a natural light Higgs mass through radiative corrections which could cause the symmetry breaking (e.g. using the Coleman-Weinberg (CW) prescription \cite{Coleman:1973jx}) 
and at the same time explain the origin of mass of the known fermions.
In this case in fact the Higgs is a fundamental particle, like in the SM, but the mechanism of symmetry breaking is completely different.
Moreover these models have a very rich phenomenology since the new scalar degrees of freedom (dof) become accessible in an energy scale that could be covered by the
second three-year LHC run and  also by the  next collider generation ---ILC (E$_{CM}$ $\lesssim$ 1TeV),  CLIC (E$_{CM}$ $\lesssim$3 TeV) or a large circular $e^+ e^-$ collider with E$_{CM}$  $\lesssim$  350 GeV and/or a $pp$ collider with E$_{CM}$  $\lesssim$ 100 TeV.

\section{The Minimal Model for an Elementary Goldstone Higgs: $SU(4) \to Sp(4)$}
\label{tree}

We will discuss here an Higgs sector embedded into a  $SU(4)\to Sp(4)$ pattern of chiral symmetry breaking  \cite{Alanne:2014kea} firstly introduced for composite dynamics in \cite{Appelquist:1999dq,Duan:2000dy,Ryttov:2008xe}.   
We  identify the Elementary Goldstone Higgs (EGH) as one of the 5 Goldstone bosons which live in the coset of the spontaneously broken global symmetry 
of the scalar sector. The latter is an enlarged symmetry group that contains the $SU_L(2)\times SU_R(2)$ (global) chiral symmetry of the SM Higgs sector. 
  In this case, 
  the most general vacuum of the theory, $E_\theta$, can be expressed as the linear combination \cite{Alanne:2014kea}
  \be E_\theta =\cos\theta \,E_B +  \sin\theta\, E_H= -E^T_\theta\,,\label{eq:E}\ee
where $0\leq\theta\leq\pi/2$ and the two independent  vacua $E_B$ and $E_H$ are defined as
\be  E_B =
\begin{pmatrix}
i\sigma_2 & 0\\
0 & -i \sigma_2
\end{pmatrix}, \quad  E_H=
\begin{pmatrix}
0 & 1\\
-1& 0
\end{pmatrix}\,.\label{vacua}
\ee
In the context of Composite (Goldstone) Higgs scenarios,  $E_B$  ($E_H$)  
is the vacuum of the theory that preserves (explicitly breaks) the EW symmetry and therefore can be used to construct  Composite Higgs (Technicolor) models 
(see \cite{Cacciapaglia:2014uja} for a detailed discussion).
\\
   The vacuum $E_\theta$ satisfies the relations
    \begin{equation}
	\label{eq:sp4algebra}
	S^a_\theta E_\theta+E_\theta\,S^{a\,T}_\theta=0,\qquad a=1,\dots,10\,,
    \end{equation}
    where $S^a_\theta$ are the 10  unbroken generators of $SU(4)$, which obey to the symplectic algebra of $Sp(4)$.  After EW symmetry breaking, the vacuum remains invariant under $U_{em}(1)$.    
The scalar sector of the theory strictly depends on the choice of the vacuum $E_\theta$. As we will see in the following, the 
alignment angle $\theta$ is completely determined by the radiative corrections and the requirement that the model reproduces the phenomenological success of the Standard Model. This framework is very different from the composite (Goldstone) Higgs scenario because in that case the different structure of the radiative corrections induced by the EW and top mass alone prefers the Technicolor limit rather than the composite Goldstone Higgs realisation.

\subsection{Scalar sector}\label{sec:scalars}

In the minimal scenario, the  scalar sector  can be constructed out of the vacuum $E_\theta$, making use of the  two-index antisymmetric irrep 
$M\sim \mathbf{6}$ of $SU(4)$,
\be \begin{split}
M   
     & =\left[\frac{1}{2} \left( \sigma + i\, \Theta\right) + \sqrt{2}\, ( \Pi_i+i \,\tilde \Pi_i) \,X^i_\theta \right] E_\theta\,,  \\ \end{split}
\label{eq:higgssector}
\ee
where  $X^i_\theta$  ($i = 1, \ldots, 5$)  are the broken generators associated to the breaking of  $SU(4)$ to $Sp(4)$.
Accordingly, the  $SU(4)$ invariant (tree-level) scalar potential with real couplings reads:
\begin{equation}
	\label{eq:pot}
	\begin{split}
	    V_M=&\frac{1}{2}m_M^2 Tr[M^{\dagger} M]+\left( c_M Pf(M)+\mathrm{h.c.}\right)+\frac{\lambda}{4}Tr [M^{\dagger}M]^2\\
	    &+\lambda_1 Tr  [M^{\dagger}MM^{\dagger}M]
		    -2\left(\lambda_2 Pf(M)^2+h.c \right)+\left(\frac{\lambda_3}{2}Tr [M^{\dagger}M] Pf(M)+h.c. \right) \,,
\end{split}\end{equation}
where $Pf(M)$ is by definition the Pfaffian of $M$, i.e.~$Pf(M)=\frac{1}{8}\, \epsilon_{i j k l} M_{ij} M_{kl}$.  Note that in absence of the terms involving  $Pf(M)$ the potential has a global $U(4)$ symmetry. 
\\
Following \cite{Gertov:2015xma}, we choose  the vacuum of the theory to be aligned in the $\sigma$ direction:

\be  \langle \sigma^2\rangle\equiv f^2= \frac{c_M-m_M^2}{4\,\lambda_{11}}\,,\qquad \lambda_{11}=\frac 1 4 (\lambda+ \lambda_1-\lambda_{2}-\lambda_3)   \label{vev-tree}\ee
where $c_M>m_{M}^2$ and  $\lambda_{11}$ is a positive effective coupling.
The tree-level scalar potential in eq.~(\ref{eq:pot}) is  independent of the parameter $\theta$ and therefore the theory at tree-level has an infinite number of degenerate vacua, of which the solution  $\theta=0$, that is $E_{0}=E_B$, preserves the EW symmetry. 
\\
We identify the fields $\Pi_{1,2,3}$ with  the longitudinal polarisation of the $W$ and $Z$  gauge bosons, whereas the EGH is given (at tree-level) by   $\Pi_4$.
Radiative corrections will provide a mass term for the Higgs boson, which in this case arises as a linear combination of
the fluctuations of the  $\sigma$ and $\Pi_4$ fields around the vacuum.
Further  the scalar fields  $\sigma$, $\Theta$ and
$\tilde{\Pi}_i$ ($i=1,\dots,5$)  acquire a non-zero mass at tree-level given by
\begin{equation}\label{treemasses}
\begin{split}
 m_{\sigma}^2 &\equiv M_\sigma^2\,\,,\, \quad m_{\Theta}^2 \equiv M_\Theta^2\,\,,\, \quad m_{\tilde\Pi_i}^2\equiv M_\Theta^2+2\lambda_f f^2
\quad \rm{with}\quad 
 \lambda_f \;\equiv \;\lambda_1 -\lambda_{2}.\end{split} \ee
\\
Finally, we notice that the $\Pi_5$ can acquire mass at tree-level by introducing a small breaking of the $SU(4)$ symmetry by adding the following operator to the potential  in eq.~(\ref{eq:pot}) 
\be V_{DM}= \frac{\mu_M^2}{8} Tr\left[ E_A M \right] Tr\left[ E_A M \right]^\ast =\frac{1}{2} \mu_M^2\left(\Pi_5^2 +\tilde \Pi_5^2 \right), 
\qquad \mbox{with} \quad 
E_A= \begin{pmatrix}
i \,\sigma_2 & 0 \\
0 & i \,\sigma_2
\end{pmatrix}\,. \label{eq:DMmass}
\ee
As shown in \cite{Alanne:2014kea}, in this case $\Pi_5$  is a stable massive particle - due to the presence of an accidental $Z_2$ symmetry -
and provides a viable Dark Matter candidate. 
Accordingly, the full tree-level scalar potential of the theory is 
\be V= V_M +V_{DM}\,. \label{VPhi}\ee
The minimum of  $V$ is still aligned in the $\sigma$ direction, but now there are new massive excitations for $\mu_M\neq 0$, that is
\begin{equation}
\begin{split}\label{treemasses2}
m_{\tilde \Pi_5}^2 & \equiv M_\Theta^2+2\lambda_f f^2+ \mu_M^2\,\,,\,    \quad m_{\Pi_5}^2 \equiv \mu_M^2\,.
\end{split}
\end{equation}
\\
All in all, once the symmetry breaking scale $f$ is fixed, the scalar sector of the theory can be described in terms of only five independent parameters: $M_\sigma$, $M_\Theta$, $\mu_M$, $\lambda_f$  and $\tilde\lambda$.

\subsection{Gauge sector}\label{sec:gauge}
We embed the EW gauge sector of the SM in $SU(4)$ so we gauge 
the $SU(2)_L\times U(1)_Y$ part of the chiral symmetry group  $SU(2)_L\times SU(2)_R \subset SU(4)$.
In this way, the  scalar degrees of freedom are minimally coupled to the EW gauge bosons via
the covariant derivative of $M$ 
	\be
	    \label{eq:covM}
	    D_{\mu}M=\partial_{\mu} M- i \left(G_{\mu}M+M \,G_{\mu}^T\right)\,, \quad \mbox{with} \quad   G_{\mu}=gW^i_{\mu}T_L^i +g' B_{\mu}T_Y\,,
	\ee
where the $SU(2)_L$ generators are $T_L^i$  ($i=1,2,3$)  and the hypercharge generator is $T_Y=T_R^3$.
The kinetic and EW gauge interaction Lagrangian of the scalar sector reads 
\be  \mathcal{L}_{gauge}= \frac{1}{2} \text{Tr}\left[D_{\mu}M^{\dagger}D^{\mu}M\right] \,, \label{eq:L}\ee
which explicitly breaks the global $SU(4)$ symmetry.
For any non vanishing $\theta$  the EW gauge group breaks spontaneously and  
the weak gauge bosons acquire non-zero masses through the Higgs-Brout-Englert mechanism that read
	\begin{equation}
	    \label{eq:WBosMasses}
	    m_W^2=\frac{1}{4}g^2f^2\sin^2\theta, \quad\text{and}\quad m_Z^2=\frac{1}{4}(g^2+g'^2)f^2\sin^2\theta\,.
	\end{equation}  
Comparing these expressions with the corresponding SM predictions 
we see that $f$ and $\theta$ must satisfy the phenomenological constraint
\be f\sin \theta\;=\;v_{\rm EW}\;\simeq\; 246 \mbox{ GeV} \,.\label{eq:ew}\ee

\subsection{Yukawa sector}\label{sec:Yukawa}
We embed each one of the SM fermion families in the fundamental irrep of $SU(4)$, namely
\be  \mathbf{L}_{\alpha}=\begin{pmatrix} L, &  \tilde \nu, & \tilde \ell
\end{pmatrix}_{\alpha L}^T\sim \mathbf{4}, \qquad  
\mathbf{Q}_{i}=\begin{pmatrix} Q, &  \tilde q^u, & \tilde q^d
\end{pmatrix}_{i \,L}^T \sim \mathbf{4},\label{eq:ql}
\ee
where $\alpha=e,\mu,\tau$ and $i=1,  2, 3$ are generation indices  and the tilde indicates the charge conjugate fields of the RH fermions, that is, for instance, 
$\tilde\nu_{\alpha L}\equiv (\nu_{\alpha R})^c$, $\tilde \ell_{\alpha L}\equiv (\ell_{\alpha R})^c$, $L_{\alpha L}\equiv (\nu_{\alpha L}, \ell_{\alpha L})^T$ and similarly for the quark fields. Notice that a RH neutrino  $\nu_{\alpha R}$ for each family must be introduced in order to define $\mathbf{L}_{\alpha}$ to transform according to the fundamental irrep of $SU(4)$.
The Yukawa couplings for the SM fermions that preserve the $SU(2)_L$ gauge symmetry can be written as:
 \begin{eqnarray}
-\mathcal{L}^\text{Yukawa} &= & \frac{Y^u_{i j}}{\sqrt{2}} \,\left(\mathbf{Q}^T_{i} \, P_a \, \mathbf{Q}_{j} \right)^\dagger Tr\left[P_a \, M\right] + 
 \frac{Y^d_{i j}}{\sqrt{2}}\, \left(\mathbf{Q}^T_{i}  \,\overline P_a\,  \mathbf{Q}_{j} \right)^\dagger Tr\left[ \overline P_a\, M\right]\nonumber\\ 
         &+& \frac{Y^\nu_{\alpha\beta}}{\sqrt{2}} \,\left(\mathbf{L}^T_{\alpha} \, P_a \, \mathbf{L}_{\beta} \right)^\dagger Tr\left[P_a \, M\right] + 
 \frac{Y^\ell_{\alpha\beta}}{\sqrt{2}}\, \left(\mathbf{L}^T_{\alpha}  \,\overline P_a\,  \mathbf{L}_{\beta} \right)^\dagger Tr\left[ \overline P_a\, M\right]\,+\,\text{h.c.}\label{LYuk1}
\end{eqnarray}
where we make use of $SU(4)$ spurion fields \cite{Galloway:2010bp} $P_a$ and $\overline{P}_a$, with  an $SU(2)_L$ index $a=1,2$ (for an explicit representation see \cite{Gertov:2015xma}).
\\
After EW symmetry breaking, we predict for the SM fermion masses
\be m_F = y_F \frac{f \sin \theta}{\sqrt{2}}\,, \ee
$y_F$ being the SM Yukawa coupling of quarks and leptons in the fermion mass basis. 
Notice, in particular, that a Dirac mass for neutrinos is generated as well. Further, 
one can implement a Type-I See-Saw adding a Majorana mass term for the RH neutrino fields, which provides an explicit breaking of the $SU(4)$ symmetry, but preserves the
EW gauge group.

\section{ Electroweak scale from radiative corrections}
\label{radiative}

A non-zero mass term for the EGH field $\Pi_4$ is generated at quantum level from those operators in the 
Lagrangian that explicitly break the global symmetry $SU(4)$, i.e. the gauge and Yukawa interactions. 
The one-loop corrections \cite{Coleman:1973jx} $\delta V(\Phi)$ of the scalar potential $V$ given in (\ref{VPhi}) takes the general expression
	    \begin{equation}
		\label{eq:deltaV}
		\delta V(\Phi)=\frac{1}{64\pi^2}\mathrm{Str}\left[{\cal M}^4 (\Phi) \left(\log\frac{{\cal M}^2(\Phi)}
		    {\mu_0^2}-C\right)\right]+V_{\mathrm{GB}},
	    \end{equation}
where in this case $\Phi\equiv(\sigma,\,\Pi_4)$ denotes  the background scalar fields that we expect to lead to the correct vacuum alignment of the theory
and  ${\cal M}(\Phi) $ is the corresponding  tree-level mass matrix. The  supertrace, $\mathrm{Str}$, is defined as
\begin{equation}
\mathrm{Str} = \sum_{\text{scalars}}-2\sum_{\text{fermions}}+3\sum_{\text{vectors}}.
\end{equation}
and we have $C = 3/2$ for scalars and fermions and $C = 5/6$ for the gauge bosons,
whereas $\mu_0$ is a reference renormalization scale. 
The terms related to the massless Goldstone bosons are described by a separate potential, $V_{GB}$, since these terms lead to infrared divergences due to their vanishing masses. There are several ways of dealing with this issue, for example adding some characteristic mass scale as an infrared regulator. However, since the massive scalars give the dominant contribution to the vacuum structure of the theory, we will simply neglect $V_{GB}$ in the following discussion.
\\
In terms of the background fields $\sigma$ and $\Pi_4$, we can write the first term in eq.~(\ref{eq:deltaV}) as
\be \delta V(\sigma,\Pi_4)\;=\; \delta V_{\mathrm{EW}}(\sigma,\Pi_4)\;+\;\delta V_{\mathrm{top}}(\sigma,\Pi_4)\;+\;\delta V_{\mathrm{sc}}(\sigma,\Pi_4), \qquad \mbox{with}\ee
\begin{align}
    \label{eq:corrEWtop}
    \begin{split}
        \delta V_{\mathrm{EW}}(\sigma,\Pi_4)\;=\;&\frac{3}{1024\pi^2}\, \phi^4\left[2g^4\left(
	    \log\frac{g^2\,\phi^2}{4\mu_0^2}-\frac{5}{6}\right)  +(g^2+g^{\prime\, 2})^2\left(\log\frac{(g^2+g^{\prime\, 2})\,\phi^2}{4\mu_0^2}
	    -\frac{5}{6}\right)\right]\,,
    \end{split}\\
        \delta V_{\mathrm{top}}(\sigma,\Pi_4)\;=\;&-\frac{3}{64\pi^2}\,\phi^4 y_t^4\left(
	    \log\frac{y_t^2\,\phi^2}{2\mu_0^2}-\frac{3}{2}\right)\,,
\end{align}
where $\phi\equiv  \sigma\sin\theta+\Pi_4\cos\theta$. We consider for simplicity only the fermion contribution in the one-loop potential
arising from the virtual top quark. Notice that both $\delta V_{\rm EW}$ and $\delta V_{\rm top}$ introduce an explicit dependence on $\theta$
in the full scalar potential of the theory.
\\
The  quantum correction originated from the scalar sector reads
\begin{equation}
\begin{split}
\delta V_{\mathrm{sc}}(\sigma,\Pi_4)\;=\;& \frac{1}{64 \pi^2} \left[-\frac{3}{2} \left(  m_{\sigma}^4 (\sigma,\Pi_4) +m_\Theta^4(\sigma,\Pi_4) +m_{\tilde\Pi_i}^4(\sigma,\Pi_4) +m_{\tilde \Pi_5}^4(\sigma,\Pi_4) +m_{\Pi_5}^4(\sigma,\Pi_4) \right)  \right. \\
& \left. + m_{\sigma}^4 (\sigma,\Pi_4) \log\left( \frac{m_{\sigma}^2 (\sigma,\Pi_4)}{\mu_0^2} \right)  +
m_{\Theta}^4 (\sigma,\Pi_4) \log\left( \frac{m_{\Theta}^2 (\sigma,\Pi_4)}{\mu_0^2} \right)  \right.\\ 
& \left. +4 m_{\tilde\Pi_i}^4 (\sigma,\Pi_4) \log\left( \frac{m_{\tilde\Pi_i}^2 (\sigma,\Pi_4)}{\mu_0^2} \right) +
m_{\tilde \Pi_5}^4 (\sigma,\Pi_4) \log\left( \frac{m_{\tilde \Pi_5}^2 (\sigma,\Pi_4)}{\mu_0^2} \right)\right. \\
& \left.+m_{\Pi_5}^4 (\sigma,\Pi_4) \log\left(\frac{m_{\Pi_5}^2 (\sigma,\Pi_4)}{\mu_0^2} \right)    \right]\,,\\
\end{split}	
\end{equation}
where the background dependent masses of the scalar fields are  
\begin{eqnarray}
\begin{split}
	m_{\sigma}^2 (\sigma,\Pi_4) &=\frac{1}{2 f^2 }M_\sigma^2 \left(3\,\sigma^2 +\Pi_4^2-f^2\right)\,, \quad 
	m_\Theta^2(\sigma,\Pi_4) = M_\Theta^2+\tilde\lambda \left(\Pi_4^2+\sigma ^2-f^2\right)\,,\\
	m_{\tilde\Pi_i}^2(\sigma,\Pi_4) &= M_\Theta^2+\tilde\lambda \left(\Pi_4^2+\sigma ^2-f^2\right)
	+2 \lambda_f \left(\Pi_4^2+\sigma ^2\right)\,,\\
	m_{\tilde \Pi_5}^2(\sigma,\Pi_4) &= m_{\Theta}^2(\sigma,\Pi_4)+\mu_M^2+2\lambda_f (\Pi_4^2+\sigma ^2)\,,\\
	m_{\Pi_5}^2(\sigma,\Pi_4) &=\frac{1}{2 f^2}M_\sigma^2 \left(\sigma^2 +\Pi_4^2-f^2\right)+\mu_M^2\,.
\end{split}\label{eq:bckfields}
\end{eqnarray}
Notice that these expressions reduce to the tree-level scalar masses (\ref{treemasses}) and (\ref{treemasses2}) 
when we evaluate them for $\langle \Phi \rangle=(f,0)$. 
\\
The minimization procedure of the full potential (tree-level plus corrections) is fully described elsewhere \cite{Alanne:2014kea} \cite{Gertov:2015xma}. Here we just want to notice that the we fix the scale $\mu_0$ in such a way that the quantum corrected potential has still an extremum in the $\sigma $ direction and only after we apply the usual minimization condition on the parameter $\theta$. Of particular importance for the determination of $\theta$ are the opposite signs of the different one loop fermionic and gauge boson contributions.

\section{A minimal phenomenological example}\label{pheno}

According to the discussion reported in the previous sections, the set of parameters that fully describes the scalar sector of the theory is the following:
$	\{f,\, \theta,\, M_\sigma,\, M_\Theta,\, \mu_M,\,\tilde \lambda,\,\lambda_f\}$.
We will discuss here a simplified scenario with mass spectrum:
\be M_\sigma\;=\;M_\Theta\;\equiv\;M_S\,,\quad \lambda_f  \;=\;0\ee
Before showing the phenomenological implications of this choice let us remark that in the model under study 
the Higgs is one of the two linear combinations of $\sigma$ and $\Pi_4$, that is 
\begin{eqnarray}
	\begin{pmatrix}
		\sigma\\
		\Pi_4
	\end{pmatrix}&=&
	\begin{pmatrix}
		\cos\alpha &-\sin\alpha\\
		\sin\alpha & \cos \alpha
	\end{pmatrix}
	\begin{pmatrix}
		H_1\\
		H_2
	\end{pmatrix}\,,\label{eq:Higgs}
\end{eqnarray}
where $H_1$ and $H_2$ are the mass eigenstates and  $\alpha$ the mixing angle,  chosen in the interval $[0,\pi/2]$. 
The observed Higgs boson will be the lightest eigenstate. 
Relevant constraints are provided by the Higgs phenomenology, starting from
the experimental value of the Higgs mass  \cite{PDG}:
 \be m_{h}=125.7\pm 0.4 \mbox{ GeV}\,.\label{Hmass}\ee
and the  SM normalised coupling strength of the Higgs with fermions and vectors:
\be
 C_F \equiv\frac{\lambda_{H_{1\left[2\right]} FF}}{\lambda_{h FF}^{SM}}=  \sin\left(\alpha+\theta\right) \left[\cos\left(\alpha+\theta\right) \right],
\qquad C_V \equiv\frac{\lambda_{H_{1\left[2\right]} VV}}{\lambda_{hVV}^{SM}} =   \sin\left(\alpha+\theta\right)\left[  \cos\left(\alpha+\theta\right)\right], \label{eq:cvcf} \ee
where $\lambda_{h FF}^{SM}\equiv y_F/\sqrt{2}$ is the SM coupling and the index between square brackets refers to $H_2$. 
The parameters $C_F$ and $C_V$ must satisfy the  experimental constraints
 \cite{CMS:2014ega}
$ C_V=1.01^{+0.07}_{-0.07}, \quad C_F=0.89 ^{+0.14}_{-0.13} \,\, \mbox{at 68\% C.L.} \label{eq:excvcf}$.
Last, we investigate also the trilinear self-coupling of $H_1$ and $H_2$ with respect to the SM prediction, 
$\lambda_{hhh}^{\rm SM}=3 \,m_h^2/v_{\rm EW}$. In this case, we have \cite{Alanne:2014kea}
\be  \frac{\lambda_{H_1 H_1 H_1}}{\lambda_{hhh}^{SM}}= v_{\rm EW}\frac{ M_\sigma^2 \cos\alpha}{f m^2_{h} }  , \qquad 
 \frac{\lambda_{H_2 H_2 H_2}}{\lambda_{hhh}^{SM}}= v_{\rm EW}\frac{ M_\sigma^2 \sin\alpha}{f m^2_{h} }\,.\label{trilinear} \ee

 \section{Numerical results and Conclusions}
In the following we  assume   $y_t=1$, a $3\,\sigma$ uncertainty on the value of the Higgs mass and use the central values of the weak gauge boson masses given in \cite{PDG}. Moreover,  we impose the perturbativity bound on the effective quartic coupling $\tilde\lambda$, i.e. $|\tilde \lambda|<4\,\pi$ and we set  the parameter $\mu_M$ to lie in the interval 
$[m_h,\,1]$~\text{TeV} with the additional constraint $\mu_M<f$. The latter ensures that $\mu_M$ introduces only a small explicit breaking of the global
$SU(4)$ symmetry.
In the minimal setup we investigate, we vary the common scalar mass  $M_S$ in the interval
\be
	m_h\;\leq\; M_S \; \leq\; 5~\text{TeV}\,.
\ee
For each random value of  $ M_S$ and $\mu_M$, we select the other parameters of the model
  imposing the experimental value of the Higgs mass and the minimisation conditions of the Coleman-Weinberg  potential. 
  In this way we extract the values of the effective quartic coupling $\tilde \lambda$ 
  and the vacuum alignment angle $\theta$,   which are, therefore, implicit functions of the dimensional parameters $M_S$ and $\mu_M$.   
  Using this procedure  we find that the mode of the distribution of the values of $\theta$ is  
 \be  \overline\theta\;=\;0.136^{+0.006}_{-0.012} \,, \label{eq:caso1theta}\ee
corresponding to $\overline{\alpha}=1.57$ and the $SU(4)$ spontaneous symmetry breaking scale of~\footnote{In the following we define the mode as the value that appears most often in a set of data. We report the error on the mode as the width evaluated at half of the mode hight. The error on the scale $f$ of the theory is computed with the standard propagation of errors.} 
\be \overline f\;=\; 1.81^{+0.08}_{-0.15}\mbox{ TeV}\,.\ee
Notice that, for a given $\theta$  the scalar mixing angle $\alpha$ is essentially  determined by imposing the experimental constraints  on $C_V$ and $C_F$.
The analysis done in \cite{Gertov:2015xma}  shows that the dynamics prefers small values of $\theta$ implying  that the Higgs boson is mostly aligned in the $\Pi_4$, the  pNGB  direction.
\\
Further, from the minimisation condition we find to a very good approximation 
 \be
 		\tilde\lambda\;\approx\; K \sin^2\theta\quad\quad\text{for}\quad\quad \sin\theta\lesssim 0.1\,,
 \ee
 where $K$ depends on $M_S$  and not on $\mu_M$ (for $M_S\approx 2.6 $ GeV, $K\approx90$).
 Henceforth, for $\sin\theta\lesssim 0.1$ the only independent parameter is the tree-level scalar mass $M_S$, which is fixed by the knowledge of the Higgs mass via  
\be m_{h}^2\approx\frac{9}{16\,\pi^2 \,v_{\rm EW}^2} \left[M_Z^4 \,\log \left( \frac{M_Z^2}{M_S^2}\right) \,+
 \,M_W^4 \,\log \left( \frac{M_W^4}{M_S^4}\right)  \,-\, v_{\rm EW}^4\left( \frac 2 3\,+\, \log \left( \frac{v_{\rm EW}^2}{2\, M_S^2}\right)\right)\right]\,.
 \label{eq:analytic}
 \ee
For $m_h = 125$ GeV and $v_{\rm EW}=246$ GeV the previous expression implies
 \be
 	M_S\;\approx\; 2.6\quad\text{TeV}\quad\text{for}\quad\sin\theta\lesssim 0.1\,.\label{boundMS}
 \ee
 We turn now to the properties of the heaviest scalar mass eigenstate defined in eq.~(\ref{eq:Higgs}), which here corresponds to  $H\equiv H_2\sim \sigma$. We find that  \cite{Gertov:2015xma}  the physical mass $M_H$ and the tree-level mass $M_S$ are close to each other once the quantum corrections are taken into account with the difference due mostly to the effects of $\mu_M$. 
 The mass of the heavy Higgs $H$ also affects the  ratio between the trilinear Higgs coupling $\lambda_{hhh}$ and the corresponding SM, see eq. \eqref{trilinear}.
 As expected from the analytic expression, there is a strong suppression for $\theta \lesssim 0.1$ corresponding to $2.6~\text{TeV}\lesssim M_H\lesssim 3$ TeV.
 \\
Finally we consider the most general possible spectrum of the theory, that is  $M_\sigma\neq M_\Theta \neq M_{\tilde \Pi_i}$. 
The parameters used in the analysis are $\tilde\lambda$, $\lambda_f$, $M_\sigma$, $M_\Theta$, $\mu_M$ and $\sin\theta$. As in the previous cases, we  generate the scalar masses and extract the values of $\theta$ and $\tilde\lambda$ that satisfy all the phenomenological constraints.
In particular,  the scalar masses are varied within the interval
\be m_h \; \leq\; M_\sigma\,,\,  M_\Theta\,, M_{\tilde \Pi_i}\; \leq\; 5\, \mbox{ TeV}\,.\ee 
We find out that a scalar mixing angle of $\overline\alpha=1.570$ and the mode of the distribution  of the $SU(4)$ breaking scale   is
\be \overline f\;=\;13.9^{+2.9}_{-2.1}\mbox{ TeV},\ee
corresponding to a mode value for the alignment angle of $ \overline\theta\;=\;0.018^{+0.004}_{-0.003}$.
 We deduce therefore that also in the most general scenario the Higgs particle is mostly  a pNGB.  
Concluding, we have shown via  a detailed analytical and numerical analyses that, a radiatively induced pNGB Higgs is possible in the SU(4)/Sp(4) context. 
The embedding of the electroweak gauge sector is parametrised by an angle  {$0\leq\theta\leq \pi/2$} which is dynamically determined to be centered around $\theta \simeq 0.02 $, corresponding to the Higgs chiral symmetry breaking scale  $ f \simeq 14~$TeV.  This is almost 60 times higher than the SM electroweak scale. Due to the perturbative nature of the theory the  new scalars remain in the few TeV energy range.

\section*{Acknowledgments}
I thank the organizers of Moriond EW 2016 for the opportunity to present this work.
 The CP$^3$-Origins center is partially supported by the DNRF:90 grant.

\end{document}